Article

# New Generation Compact Linear Accelerator for Low-Current, Low-Energy Multiple Applications

Jorge Feuchtwanger [1,2,*], Victor Etxebarria [1], Joaquin Portilla [1], Josu Jugo [1], Iñigo Arredondo [1], Inari Badillo [1], Estibaliz Asua [1], Nicolas Vallis [1], Mikel Elorza [1], Beñat Alberdi [3], Rafael Enparantza [4], Iratxe Ariz [4], Iñigo Muñoz [4], Unai Etxebeste [5] and Iñaki Hernandez [5]

1   IZPILab-Beam Laboratory, Faculty of Science and Technology, University of the Basque Country, UPV/EHU, 48940 Leioa, Spain; victor.etxebarria@ehu.eus (V.E.); joaquin.portilla@ehu.eus (J.P.); josu.jugo@ehu.eus (J.J.); inigo.arredondo@ehu.eus (I.A.); inari.badillo@ehu.eus (I.B.); estibaliz.asua@ehu.eus (E.A.); nicolas.vallis@ehu.eus (N.V.); mikel.elorza@ehu.eus (M.E.)
2   Ikerbasque, Basque Foundation for Science, 48009 Bilbao, Spain
3   Helmholtz-Zentrum Berlin (HZB), Albert-Einstein-Strasse 15, 14109 Berlin, Germany; benat.alberdi_esuain@helmholtz-berlin.de
4   TEKNIKER, Basque Research and Technology Alliance (BRTA), 20600 Eibar, Spain; rafael.enparantza@tekniker.es (R.E.); iratxe.ariz@tekniker.es (I.A.); inigo.munoz@tekniker.es (I.M.)
5   Egile S.L., 20850 Mendaro, Spain; unai.etxebeste@egile.es (U.E.); inaki.hernandez@egile.es (I.H.)
*   Correspondence: jorge.feuchtwangerm@ehu.eus

**Abstract:** A new compact linear proton accelerator project (named LINAC 7) for multiple low-current applications, designed and built in-house at the Beam Laboratory of the University of the Basque Country (UPV/EHU) is described. The project combines the University, a research technology center and a private company with the aim of designing and building a compact, low-current proton accelerator capable of accelerating particles up to 7 MeV. In this paper, we present an overview of the accelerator design, summarize the progress and testing of the components that have been built, and describe the components that are being designed that will allow us to achieve the final desired energy of 7 MeV.

## 1. Introduction

New compact particle accelerators are being developed around many kinds of state-of-the-art technologies and applications in such a way that near future industrial elements are built through a new generation of remarkable important enhanced components custom-designed for these newest accelerators [1]. Ion linear accelerators (LINACs) can be a valid alternative to many important low-current, low-energy applications, including materials science analysis [2,3], doping semiconductors or polymer modification [4,5], as well as in medical bioapplications [6,7], among other fields. Our LINAC 7 project consists of a new compact linear proton accelerator developed through the IZPILab-Beam Laboratory at the University of the Basque Country (UPV/EHU), whose first components are already operational [8,9]. The final 7 MeV energy was chosen because it can be achieved with normally conducting elements and is high enough to have several practical applications. For example, it is high enough to produce neutrons if a lithium target is used. Moreover, LINAC 7 can also be used to produce short-lived radio isotopes for medical applications, and, finally, it can be used as the injector to other structures if higher energies are required.

The accelerator is composed of an electron cyclotron resonance (ECR) ion source, a low energy beam transport (LEBT), a radio frequency quadrupole (RFQ), a medium energy beam transport (MEBT), a drift tube linac (DTL), and, finally, a beam stop. In Figure 1, a simplified schematic of the accelerator with the energies achieved after each element can be seen. The boxes with solid borders indicate the components that have been completed and are at different stages of testing. Dotted borders indicate components that are in the design phase.

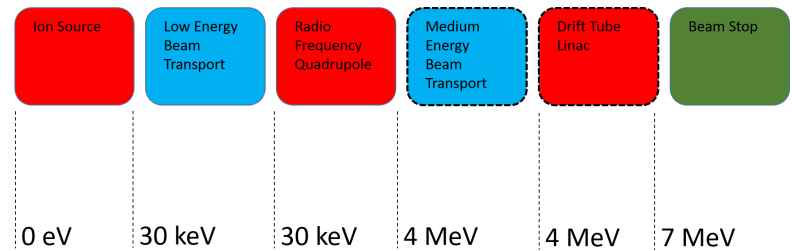

**Figure 1.** Simplified schematic of LINAC 7.

The different research groups that are taking part in this project came together because they provide complementary experiences that, as a whole, are needed to complete the project. The University of the Basque Country (UPV/EHU) provides the theoretical knowledge that serves as the basis for the design, and is also responsible for the operation and testing of the components. TEKNIKER is a large research foundation that is very experienced in the design for fabrication of components and have been responsible for the delivery of large components to European research facilities, such as RAL, STFC, and ILL, among others. The last partner in this group is Egile, a company specializing in high precision machining, and which has considerable experience in manufacturing parts for large accelerator facilities, such as CERN.

## 2. LINAC 7 Accelerator Component Descriptions

### 2.1. Ion Source

The first component in the accelerator is an off-resonance electron cyclotron resonance (ECR) ion source. It consists of a plasma chamber made from a special length CF 63 vacuum pipe. One end of the tube is closed with a special cap that includes a coaxial RF feedthrough and a 3 mm tube for the gas feed. The other end of the cylinder is partially closed off by the plasma electrode. The ECR magnetic field is provided by a Hallbach-type permanent magnet arrangement of eight bars that generates a field coaxial with the chamber. Both the magnitude of the magnetic field, as well as the chamber dimensions, have been set to correspond to a 3 GHz RF power input. The RF power is supplied to the chamber through a DC break designed to allow RF to be transmitted, while isolating the RF amplifier from the high voltage used for extraction. The chamber is connected to the rest of the system through an alumina isolator. The setup can be seen in Figure 2.

The extraction optics are supported by a special flange with a high voltage feedthrough on the ground side of the alumina isolator, and consists of a set of three electrodes that act as an Einzel lens. The two electrodes on the ends are connected to the ground, while the central electrode is connected to a resistive divider that always keeps it at half of the extraction voltage, avoiding the need for an extra high voltage supply. The proton optics are shown in Figure 3. This was set up in this way to minimize the cost of the ion source by having no instruments that require power or communications at high voltage. The ion source is designed to produce 30 keV protons.

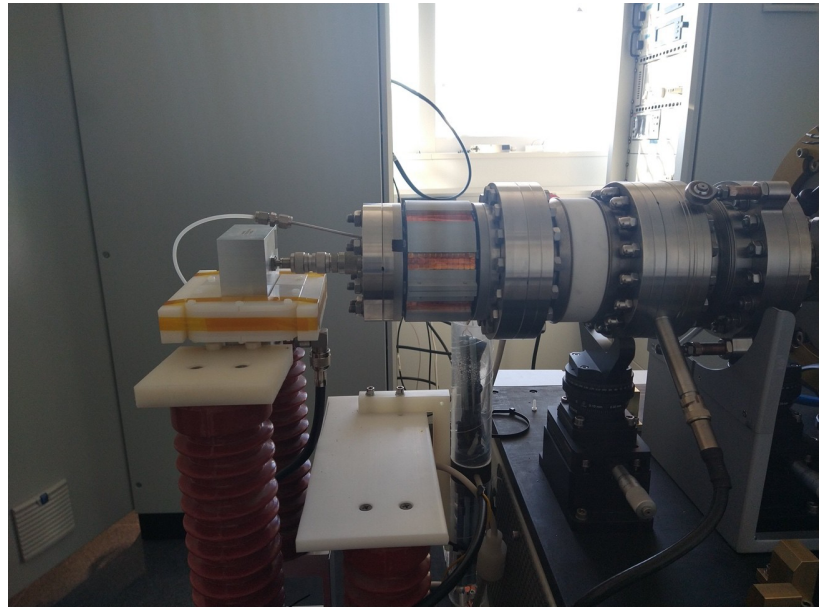

**Figure 2.** Off-resonance ECR ion source for LINAC 7.

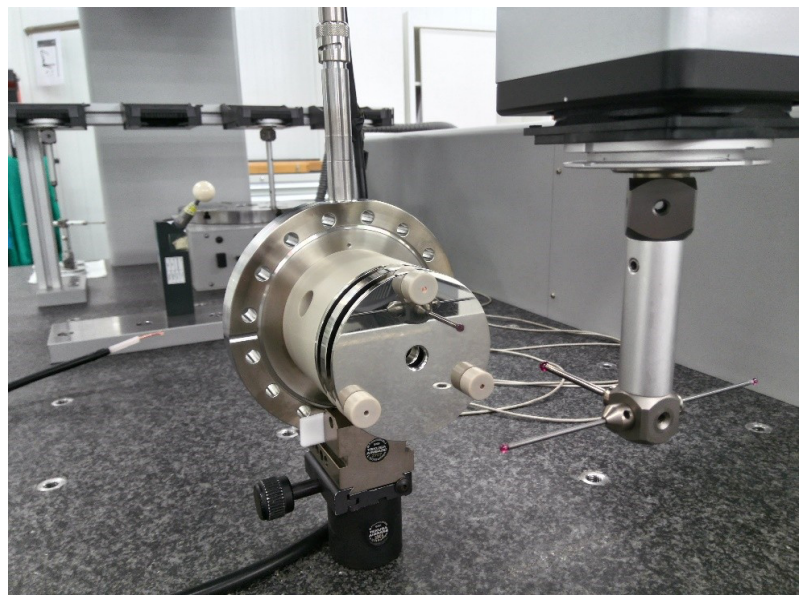

**Figure 3.** Einzel lens for the ECR ion source for LINAC 7 being measured after final assembly using a Zeiss coordinate measuring machine at TEKNIKER.

*2.2. Low Energy Beam Transport*

In order to match the beam that the ion source generates to the input requirements of the radiofrequency quadrupole (RFQ), a low energy beam transport (LEBT) is used. Like the ion source, this element is built and in use. It consists of two solenoids that are used to transform the divergent beam form from the source into the convergent one that the RFQ requires. The vacuum vessel consists of a custom-made six-way cross. Four limbs that are in the same plane have a nominal diameter of 100 mm, and have CF flanges on the ends. The other two limbs are of a 40 mm nominal diameter and have quickCF flanges on the ends. The DN 40 limbs are used as the beam pipe, and a solenoid fits over each one of them. Of the DN 100 branches, two are used for beam diagnostics, one for the turbomolecular pump that generates the vacuum for the ion source and the LEBT, and the last for a vacuum gauge and a vacuum switch. The entire setup can be seen in Figure 4; the left side shows a photograph of the system and the right side a cross-section of the CAD model with the elements labeled.

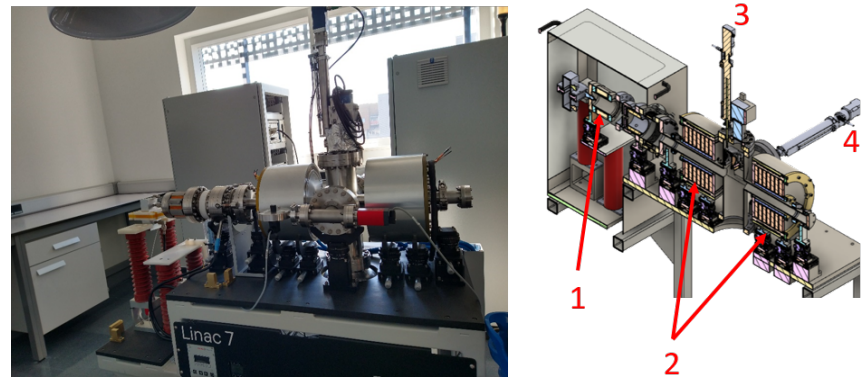

**Figure 4.** On the (**left**), a photograph of the LEBT as built. On the (**right**), a cross-section of the 3D CAD model, 1 is the ion source, 2 is the solenoids, 3 is the pepperpot and 4 is the Faraday cup.

A Faraday cup and a pepperpot are used as beam diagnostics in the LEBT—both were custom designed and built for this LEBT. Their size is such that they can capture the entire beam when deployed and fit entirely inside their respective DN 100 branches when retracted so that they do not interfere with the beam during normal operation. Standard motorized vacuum positioners are used to position the diagnostics.

*2.3. Radio Frequency Quadrupole (RFQ)*

The particle beam generated in the source will be accelerated using a radio frequency quadrupole. This element is still in the design phase. In order to reduce the size of the RFQ its operating frequency was chosen to be 750 MHz. To further reduce the length, the choice was made to dispense with the gentle buncher section of the RFQ. This comes at the cost of a lower particle transmission, but this is not an issue since the goal is a low current accelerator. The particles will be lost at the entrance of the RFQ, meaning they will not have enough energy to generate any activation of the copper. The final RFQ is expected to achieve an output energy of 4 MeV at the end of 1.5 m, and capture between 15% and 30% of the particles at the entrance. The input parameters of the RFQ were set by the measured values of the beam in the LEBT. The design of the cavity has been completed and a cold model was manufactured by Egile out of ETP copper (electrolytic tough pitch copper, UNS C11000) to validate the design process, as well as the ability to machine the RFQ with the required precision, that is, tolerances lower than 0.1 mm. Figure 5 shows the RFQ cold model under measurement. ETP copper was used because the cold model is not intended for a vacuum, but it still provides information about the achievable tolerances and surface finishes during machining, and has roughly the same conductivity as oxygen-free high-conductivity copper for the electromagnetic testing.

A key part of the machining process of the RFQ is the modulation. For this reason the cold model includes a dummy modulation that is representative of what can be expected in the final RFQ. Figure 6 shows one of the minor segments of the RFQ cold model where the dummy modulation is clearly visible.

Geometric tolerances were verified using a Zeiss CMM (coordinate measuring machine, Carl Zeiss Iberia, S.L.—Division Metrologia Industrial, Tres Cantos, España) at TEKNIKER. All general dimensions were found to be within less than 0.1 mm of the design values and below 0.05 mm in the modulation. Design of the final modulation is in process; the main goal is to achieve the highest acceleration in the shortest distance at the expense of particle transmission.

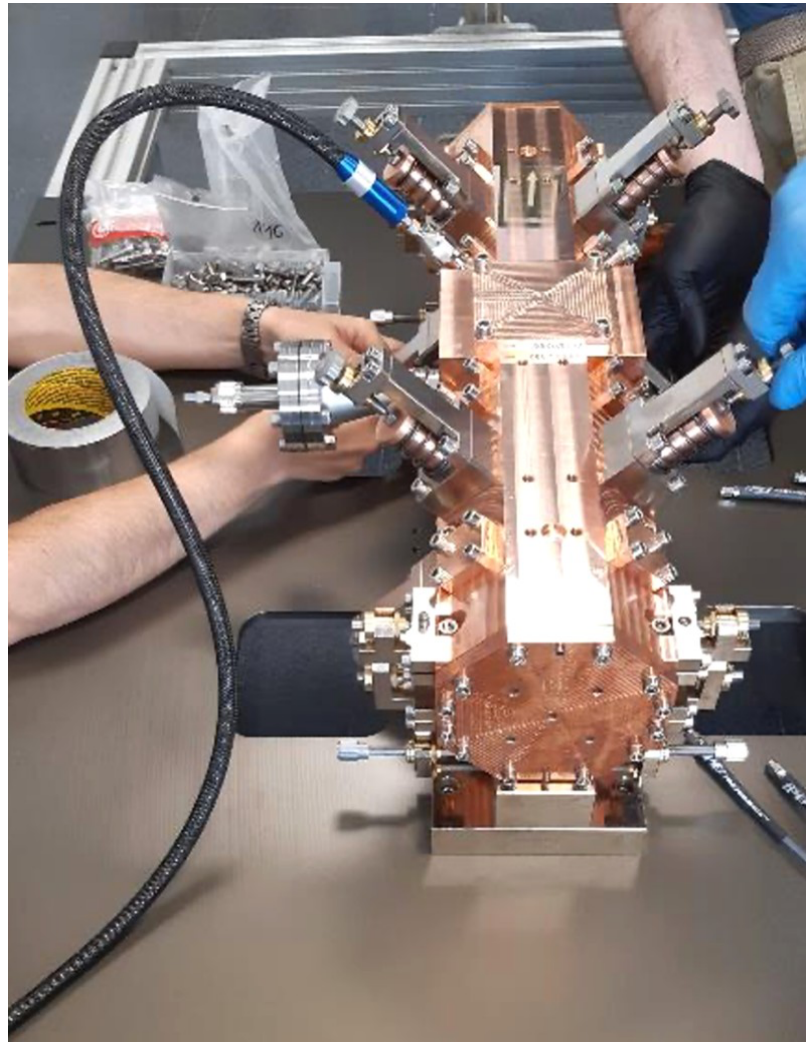

**Figure 5.** RFQ cold model being installed for RF characterization tuning and electromagnetic field measurement.

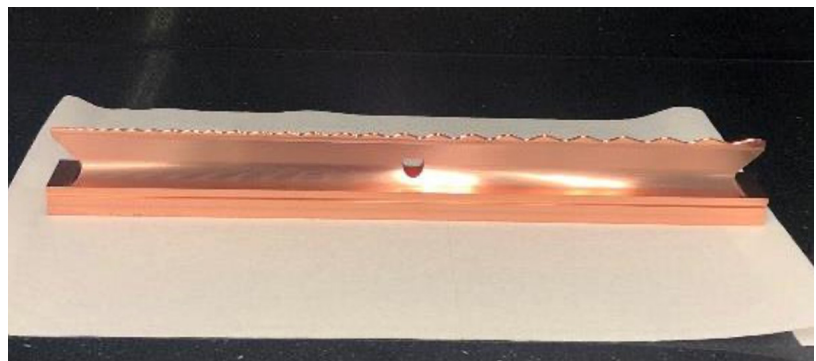

**Figure 6.** Minor vane inside the RFQ cold model clearly showing the dummy modulation.

*2.4. Medium Energy Beam Transport (MEBT)*

Once the design of the RFQ is finalized the output parameters will be analyzed, and, based on them, a decision will be made on whether an MEBT is needed or not, before passing the beam on to the next accelerating structure. If an MEBT is needed, the topology will be very conventional—it will consist of pill-box type bunching cavities, magnetic quadruples and will include button-type beam position monitors to check the position of the beam in the pipe and to measure the current.

### 2.5. Drift Tube Linac (DTL)

The final energy desired for the accelerator is 7 MeV as this is sufficient to produce radio chemicals for medical diagnostics, and, if impacted against a lithium target, it can be used to produce a low flow of neutrons. This is important for the design process of the DTL because these applications only require the right energy and not as high a quality beam as one that is to be further accelerated. The design of the DTL will be concurrent with the decision on the need of the MEBT, with the goal of finding the lowest cost solution that minimizes particle loss because, unlike at the beginning of the accelerator, here particle loss will result in unwanted radiation production. The DTL is a standing wave structure devoted to accelerate particles based on the Alvarez drift tube. The idea was proposed a long time ago but remains in use in most proton and other ion linacs. The resulting standing wave is employed to accelerate the particles. Drift tubes are suspended by stems and the beam passes through them at the locations at which the oscillating standing wave would produce beam deceleration. Drift tubes include focusing magnets which help to avoid beam spatial dispersion. In order to reduce the number of power supplies needed, the proposed DTL will use permanent magnet quadruples in the drift tubes.

### 2.6. Beam Stop

A beam stop is a crucial element for the commissioning and testing of the accelerator. For this reason, it was decided to move the design and fabrication of the beam stop to the beginning of the project. Rather than choosing a simple beam dump, it was decided to design a Faraday cup capable of continuously withstanding the final beam current and energy to be used as the beam dump. This element will be used at every stage of the project and will be moved to the back of every new element that is installed at the end of the beam line.

Figure 7 shows a photograph of the completed beam stop during vacuum testing installed at the end of the LEBT. Graphite was chosen for the actual beam strike surface for several reasons. It is a conductive material with a very high melting point, but, most importantly, as is reported in the literature, no neutrons are produced when a 7 MeV proton beam impacts against it. To minimize the energy flux on the surface, the beam is impacted at a shallow angle against the graphite. The energy deposition on the graphite was simulated using GEANT4 [10], and the resulting power deposition map was used to simulate the heat generation in Ansys [11]. It was found that the beam stop can function without water cooling when placed after the ion source, but has to be water-cooled when used for the full 7 MeV beam. However, as a safety measure, the beam stop was designed so that it could operate for up to 5 min without the external surface of the vacuum enclosure exceeding 300 degrees Celsius. Two thermocouple wells were drilled into the copper piece that supports the graphite, and the temperature measurements will be used as inputs to the safety system. Should there be an interruption in the cooling water supply the temperature rise will trigger a shutdown. The overdesign of the beam stop provides enough time for this to be completed safely before the temperature increase could cause any damage.

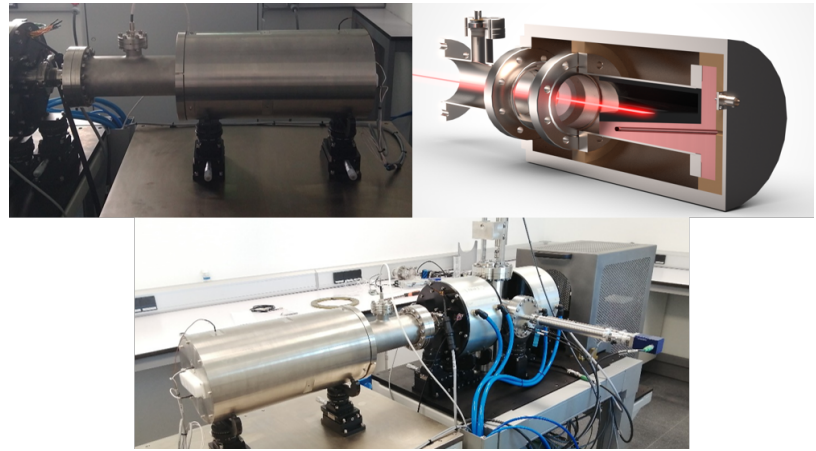

**Figure 7.** LINAC 7 beam stop (Faraday cup). Installed beam stop at the end of the LEBT for initial testing under low energy. Top right shows a cross-section of the cad model; the slanted impact face is visible there.

*2.7. Control Systems*

Each of the elements of the accelerator requires a set of specific auxiliary systems to function, for example, as vacuum, refrigeration, high voltage power supply, magnet power supply, or RF power. In order to maintain the modularity of the accelerator, each accelerator element has its own control cabinets that house the controllers for the auxiliary systems. All these cabinets are connected to a general control system.

The architecture of the control system is designed following a modular scheme. Each main element in the linac has a dedicated controller implemented in PXI or CompactRIO hardware and these controllers are connected to each other over a TCP/IP network. The auxiliary systems and diagnostics connect to their assigned controller directly over a TCP/IP connection. The PXIs or cRIOs have local controllers running in real-time LabVIEW and accept commands from the supervision and monitoring system. The local controllers acquire the relevant data from the machine and manage the local auxiliary elements and diagnostics, allowing, for example, for the insertion or retraction of the Faraday cup or pepperpot in the LEBT. The high power RF to the cavities, such as the RFQ,will be controlled by a low-level RF control system, which will use a direct digital architecture following the work presented in [12]. The control cabinets also include the interlocks and safety control system, which is implemented using PILZ safety PLC hardware. The controllers for each element connect to the Pilz hardware that insures the safety of the installation, especially the parts that could pose a danger to people, for example, unlatching the high-voltage enclosure for the ion source. All data is automatically sent to a monitoring system and selected parameters and measurements are saved in an InFluxDB database. Because this control architecture is modular, the scalability of the system is assured, allowing the integration of other technologies following a structure similar to an EPICS network that, in the future, could be integrated into an actual EPICS network.

## 3. Experiment and Simulations

*3.1. Ion Source*

The described ECR ion source is working as required for LINAC 7. As an illustration of the extracted proton beam, Figure 8 shows an image measured for a simple 6 keV beam spot on a P43 phosphorescent screen when focused to the smallest size achievable.

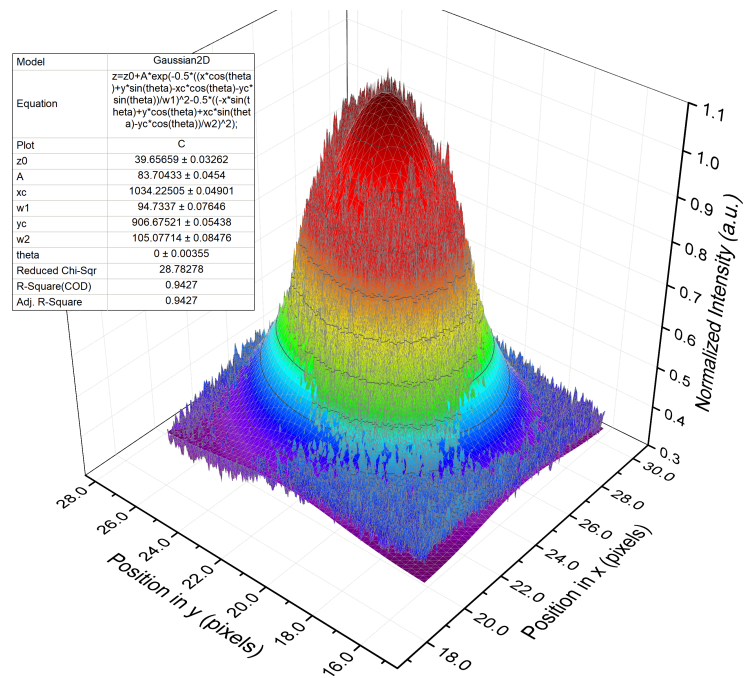

**Figure 8.** Image intensity profile measured from the image generated on the P43 screen for very low energy proton beam, and the Gaussian that fits it.

The ion source has been shown to be capable of generating beam currents of up to 10 uA with emittances $e_x$ = 0.0099543 mm mrad and $e_y$ = 0.0122865 mm mrad, measured with a pepperpot [13]. These measurements are shown in Figure 9.

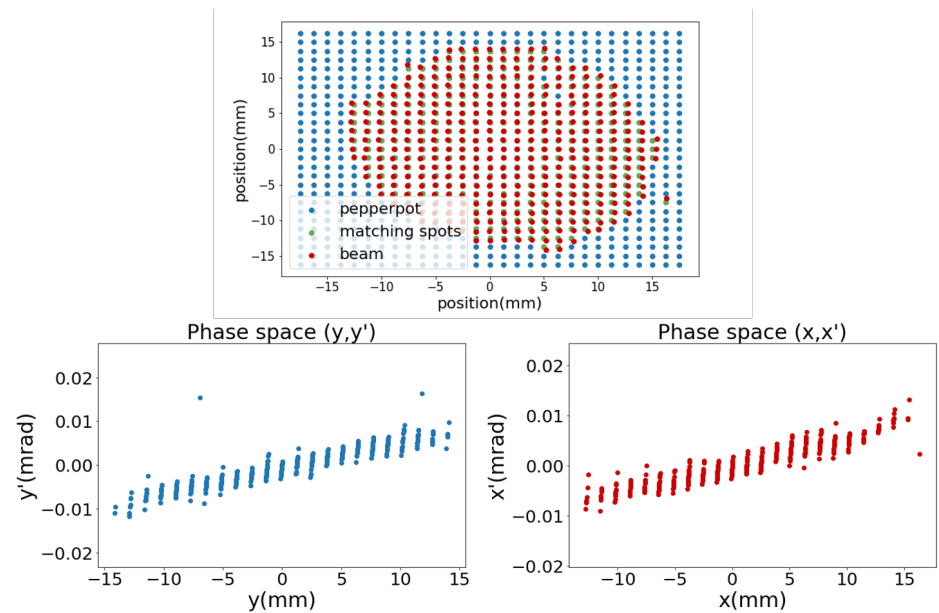

**Figure 9.** Beam emittance as measured with the pepperpot located in the central section of the LEBT.

### 3.2. Low-Energy Beam Transport

The LEBT is working as designed in conjunction with the ion source, through the double proton optics combination comprising the Einzel electric field lens and the solenoids magnetic field lens, shown in Figure 10, to transport the proton beam from the ion source to the RFQ.

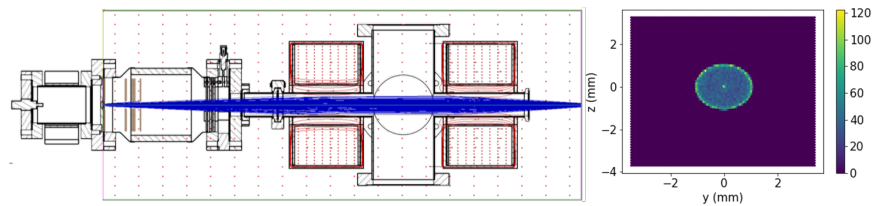

**Figure 10.** On the left, simulation of the beam (in blue) first focused at the extraction from the ion source, and then at the two solenoids in the LEBT. On the right, the particle distribution in the X-Y plane for a 3000 particle simulation.

*3.3. Radio Frequency Quadrupole*

A high-voltage has to be produced between the RFQ vanes by exciting a quadrupolar mode in such a way that an extremely effective electrostatic lens effect is produced over the particle beam travelling along the axis of the RFQ. The modulation of the vane electrodes shapes along the structure enables particle acceleration together with the focusing and bunching effect by means of a single RF quadrupolar field at the resonant frequency. A continuous, alternating quadrupole channel is formed along the RFQ axis with periodic length. The unitary length depends on the wavelength and the increasing energy of the particles along the RFQ. The RFQ operates under vacuum conditions to avoid collisions between air molecules and the particle beam. The goals of the EM design are to provide the desired RF field structure at the working frequency, which consists of a TE210 mode with minimal losses, by keeping away in frequency other undesired modes that can appear close to the desired mode due to the complex geometry of the RFQ. The EM design also has to consider the signal injection, tuning elements, and the effect of other elements, such as vacuum ports, signal pick-ups, etc. Typical Q factors achieved are of the order of several thousand, but the operation of an RFQ needs a high amount of RF power, so thermal issues have to be managed in order to avoid undesired effects. Thermal simulations are needed to optimize the RFQ, both from the EM and the mechanical points of view. Water channels are commonly implemented through the vanes along the structure to dissipate thermal power in the vane tips, whose shape is critical to the RFQ operation. In Figure 11, the proposed transversal section and the quadrupolar electrical fields can be observed. The corresponding magnetic fields are shown in Figure 12.

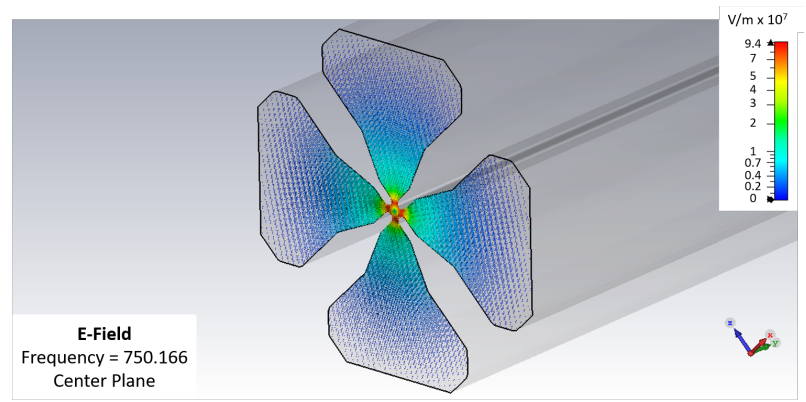

**Figure 11.** Quadrupolar electric fields in a transversal cut of the RFQ structure.

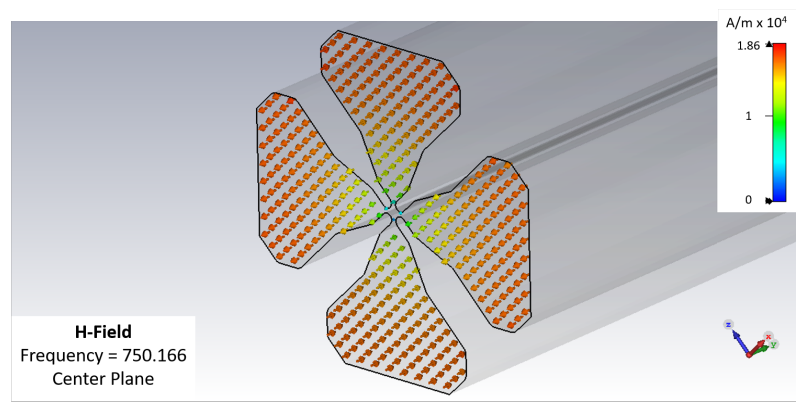

**Figure 12.** Quadrupolar magnetic fields in a transversal cut of the RFQ structure.

## 4. Conclusions

A new compact proton Linac project (LINAC 7), capable of accelerating the beam up to 7 MeV in energy and 10 uA in current, has been described and demonstrated. Newly designed components of the machine, including its ion source, the low-energy beam transport (LEBT), the radio frequency quadrupole (RFQ), and its beam stop have already been designed, built and tested at the IZPILab-Beam Laboratory of the University of the Basque Country (UPV/EHU). The ion source has been measured to provide an initial 30 keV proton beam extracted with excellent rms emittance around 0.01 mm mrad. It has been demonstrated that the former proton beam can be effectively guided through the proposed LEBT to the accelerating RFQ cavity, the capability of which for conveying several MeV in energy to the beam is also shown. Some of the applications for which this LINAC 7 could be used include industrial, scientific and medical applications, for example, producing neutrons with a lithium target, generating proton-induced X rays for materials science, doping semiconductors, or producing short-lived radio isotopes for medical applications, as well as use as an injector to other structures if higher energies are required.

**Author Contributions:** J.F., J.P., B.A. and N.V. undertook the design and multiphysics simulation of the components; J.J., I.A. (Iñgo Arredondo) and, I.B. designed and implemented the control system; I.A. (Iratxe Ariz), and I.M. performed the detail engineering and thermal analysis; U.E. and I.H. contributed to the detail design and were responsible for manufacturing the components; E.A., J.F., I.B., M.E. and I.A. (Iñgo Arredondo) carried out the experimental measurements; R.E. and V.E. coordinated the work. All authors have read and agreed to the published version of the manuscript.

**Funding:** This research was funded by the Basque Government, Department of Economic Development, Sustainability and Environment under codes Elkartek KK-2020/00003 and KK-2021/00029, and by the University of the Basque Country UPV/EHU Research Group ref. GIU18/196.

**Institutional Review Board Statement:** Not applicable for studies not involving humans or animals.

**Informed Consent Statement:** Not applicable.

**Data Availability Statement:** Not applicable.

**Conflicts of Interest:** The authors declare no conflict of interest.